\documentclass[fleqn,usenatbib]{mnras}

\usepackage{newtxtext,newtxmath}
\usepackage[T1]{fontenc}
\usepackage{ae,aecompl}

\usepackage{graphicx}	% Including figure files
\usepackage{amsmath}	% Advanced maths commands
\usepackage{amssymb}	% Extra maths symbols
\usepackage[dvipsnames]{xcolor}

\title[rSFMS \& the clumpiness of star formation]{Variations in the slope of the resolved star-forming main sequence: a tool for constraining the mass of star-forming regions} 

\author[M. H. Hani et al.]{
  \parbox[t]{\textwidth}{
  {Maan H. Hani$^{1}$\thanks{E-mail: mhani@uvic.ca}\thanks{Vanier Scholar},}
  {Christopher C. Hayward$^{2}$,}
  {Matthew E. Orr$^{3}$,}
  {Sara L. Ellison$^{1}$,}
  {Paul Torrey$^{4}$,}
  {Norm Murray$^{5}$,}
  {Andrew Wetzel$^{6}$,}
  {and Claude-Andr\'e Faucher-Gigu\`ere$^{7}$}
  }
\\\\
$^{1}$Department of Physics and Astronomy, University of Victoria, Victoria, British Columbia, V8P 1A1, Canada\\
$^{2}$Center for Computational Astrophysics, Flatiron Institute, 162 Fifth Avenue, New York, NY 10010, USA\\
$^{3}$TAPIR, Mailcode 350-17, California Institute of Technology, Pasadena, CA 91125, USA\\
$^{4}$Department of Astronomy, University of Florida, 211 Bryant Space Sciences Center, Gainesville, FL 32611, USA\\
$^{5}$Canadian Institute for Theoretical Astrophysics, 60 St George Street, University of Toronto, ON M5S 3H8, Canada\\
$^{6}$Department of Physics, University of California, Davis, CA 95616, USA\\
$^{7}$Department of Physics and Astronomy and CIERA, Northwestern University, 2145 Sheridan Road, Evanston, IL 60208, USA\\
}

\date{Accepted XXX. Received YYY; in original form ZZZ}

\pubyear{2019}

\begin{document}
\label{firstpage}
\pagerange{\pageref{firstpage}--\pageref{lastpage}}
\maketitle

%%%%%%%%%%%%%%%%%%%%%%%%%%%%%%%%%%%%%%%%%%%%%%%%%%

%%%%%%%%%%%%%%%%% BODY OF PAPER %%%%%%%%%%%%%%%%%%

\begin{abstract}
The correlation between galaxies' integrated stellar masses and star formation rates (the `star formation main sequence'; SFMS) is a well-established scaling relation. Recently, surveys have found a relationship between the star formation rate and stellar mass surface densities on kpc and sub-kpc scales (the `resolved SFMS'; rSFMS). In this work, we demonstrate that the rSFMS emerges naturally in FIRE-2 zoom-in simulations of Milky Way-mass galaxies.
We make SFR and stellar mass maps of the simulated galaxies at a variety of spatial resolutions and star formation averaging time-scales and fit the rSFMS using multiple methods from the literature.
While the absolute value of the SFMS slope ($\alpha_\mathrm{MS}$) depends on the fitting method,
the slope is steeper for longer star formation time-scales and lower spatial resolutions regardless of the fitting method employed.
We present a toy model that quantitatively captures the dependence of the simulated galaxies' $\alpha_\mathrm{MS}$ on spatial resolution
and use it to illustrate how this dependence can be used to constrain the characteristic mass of star-forming clumps. 
\end{abstract}


\section{Introduction}
\label{sec:intro}
\noindent 
Over the past decade, the `star formation main sequence' \citep[SFMS; ][]{2004MNRAS.351.1151B, 2007ApJ...670..156D, 2007ApJ...660L..43N, 2007ApJS..173..267S}, the roughly linear correlation between actively star-forming galaxies' total star formation rates (SFRs) and stellar masses, has become a key scaling relation in observational galaxy formation. A wealth of observations show that this relatively tight correlation \citep[with an intrinsic scatter of $\la 0.3$ dex; ][]{2014ApJS..214...15S, 2015ApJ...811L..12W} holds across many orders of magnitude in stellar mass and at all redshifts probed \citep[$0\le z \le 5$; e.g. ][]{2004MNRAS.351.1151B, 2007ApJ...670..156D, 2007A&A...468...33E, 2007ApJ...660L..43N, 2009ApJ...698L.116P, 2015A&A...575A..74S},  although quantitative details such as the precise values and redshift evolution of the normalization, power-law index, and scatter are still debated \citep[e.g. ][]{2014ApJS..214...15S, 2015ApJ...811L..12W, 2015A&A...575A..74S}. This correlation holds in effectively all modern galaxy formation simulations \citep[e.g. ][]{2014MNRAS.438.1985T, 2015MNRAS.450.4486F, 2015MNRAS.447.3548S, 2017MNRAS.466...88S} and semi-analytic models \citep[e.g.][]{2010MNRAS.405.1690D, 2016MNRAS.461.1760H, 2017MNRAS.465..619B}. 

In recent years, various authors have studied the spatially resolved SFR--stellar mass relation (rSFMS) using photometric or integral field unit (IFU) observations of individual galaxies \citep[e.g. ][]{2016ApJ...821L..26C, 2016A&A...590A..44G, 2017MNRAS.469.2806A, 2017ApJ...851L..24H, 2017MNRAS.466.1192M, 2018MNRAS.474.2039E, 2018ApJ...857...17L, 2019MNRAS.484.5009E}. While the observed rSFMS on $\sim$kpc and $100$ pc scales has been documented by recent observational surveys, the correlation between the resolved stellar mass surface density ($\Sigma_\star$) and SFR surface density ($\Sigma_\mathrm{SFR}$) has not been explored in depth in numerical simulations. To date, \citet{2019MNRAS.485.5715T} remains the only study to investigate the existence and redshift evolution of the rSFMS in numerical simulations, particularly the EAGLE cosmological simulation \citep{2015MNRAS.450.1937C, 2015MNRAS.446..521S}.

We present a study of the rSFMS in cosmological zoom-in simulations that resolve the dense ISM and include explicit stellar feedback. Specifically, we analyse a set of Milky Way-mass ($M_{\rm halo} \sim 10^{12} {\rm M}_{\odot}$ at $z\approx 0$) simulated galaxies from the Feedback in Realistic Environments (FIRE) Project\footnote{\url{http://www.fire.northwestern.edu}} run using the FIRE-2 physics modules \citep{2018MNRAS.480..800H}. For each galaxy, we generate SFR and stellar maps at different spatial resolutions using multiple commonly used SFR averaging time-scales (10 Myr and 100 Myr) to crudely investigate the sensitivity of our results to the SFR tracer. We find that the slope of the rSFMS depends on both the spatial resolution of the maps and the time-scale of the SFR tracer. We then present a toy model to interpret the simulations and to demonstrate how the dependence of the slope of the rSFMS on spatial resolution and SFR tracer can be used to constrain the characteristic mass (or, equivalently, area filling factor) of star-forming clumps when individual clumps are not resolved.

\section{Simulations and Analysis Methods}
\label{sec:methods}

% --------------------------------------------------------------------------------
% FIRE-2 simulations
% --------------------------------------------------------------------------------
In this Letter, we examine the existence and robustness (to observational resolution, SFR tracer time-scale, and fitting method) of the resolved $\Sigma_\star-\Sigma_\mathrm{SFR}$ correlation using simulations of Milky Way-mass galaxies from the FIRE project; all details of the methods are described in Section 2 of \citet{2018MNRAS.480..800H}. The simulations were run using \textsc{gizmo}\footnote{\url{http://www.tapir.caltech.edu/~phopkins/Site/GIZMO.html}} \citep{2015MNRAS.450...53H} in its meshless finite-mass (MFM) configuration.
Both hydrodynamic and gravitational (force-softening) spatial resolution are set in a fully adaptive Lagrangian manner; mass resolution is fixed. The simulations include cooling and heating from a meta-galactic background and local stellar sources from $T\sim10-10^{10}\,$K; star formation in locally self-gravitating, dense, self-shielding molecular, Jeans-unstable gas, assuming an instantaneous efficiency of 100 per cent per local free-fall time ($\rho_{\rm SFR} = \rho_{\rm mol}/t_{\rm ff}$); and stellar feedback from OB \&\ AGB mass loss, SNe Ia \&\ II, and multi-wavelength photo-heating and radiation pressure, with inputs taken directly from stellar evolution models. The FIRE physics, source code, and all numerical parameters are {\em exactly} identical to those in \citet{2018MNRAS.480..800H}.

% FIRE galaxies et resolution
The work presented here uses isolated Milky Way-mass FIRE-2 galaxies with $z=0$ halo masses ranging between $1.08 \times 10^{12}$ and $1.71 \times 10^{12}$ M$_\odot$. In particular, we use the following FIRE-2 galaxies: \texttt{m12b}, \texttt{m12c}, \texttt{m12f}, \texttt{m12i}, \texttt{m12m}, \texttt{m12r}, and \texttt{m12w}. The halos were simulated with a baryon mass resolution $m_\mathrm{b} = 7,100$ M$_\odot$. We include all snapshots with $z \lesssim0.1$ in our analysis (70 snapshots). For more details about the galaxies used, we refer the reader to \citet{2016ApJ...827L..23W}, %\citet{2017MNRAS.471.1709G}, \citet{2018MNRAS.481.4133G}, 
\citet{2018MNRAS.480..800H}, and \citet{2019arXiv190411508S}.

\begin{figure}
\centering
\includegraphics[width=0.65\columnwidth]{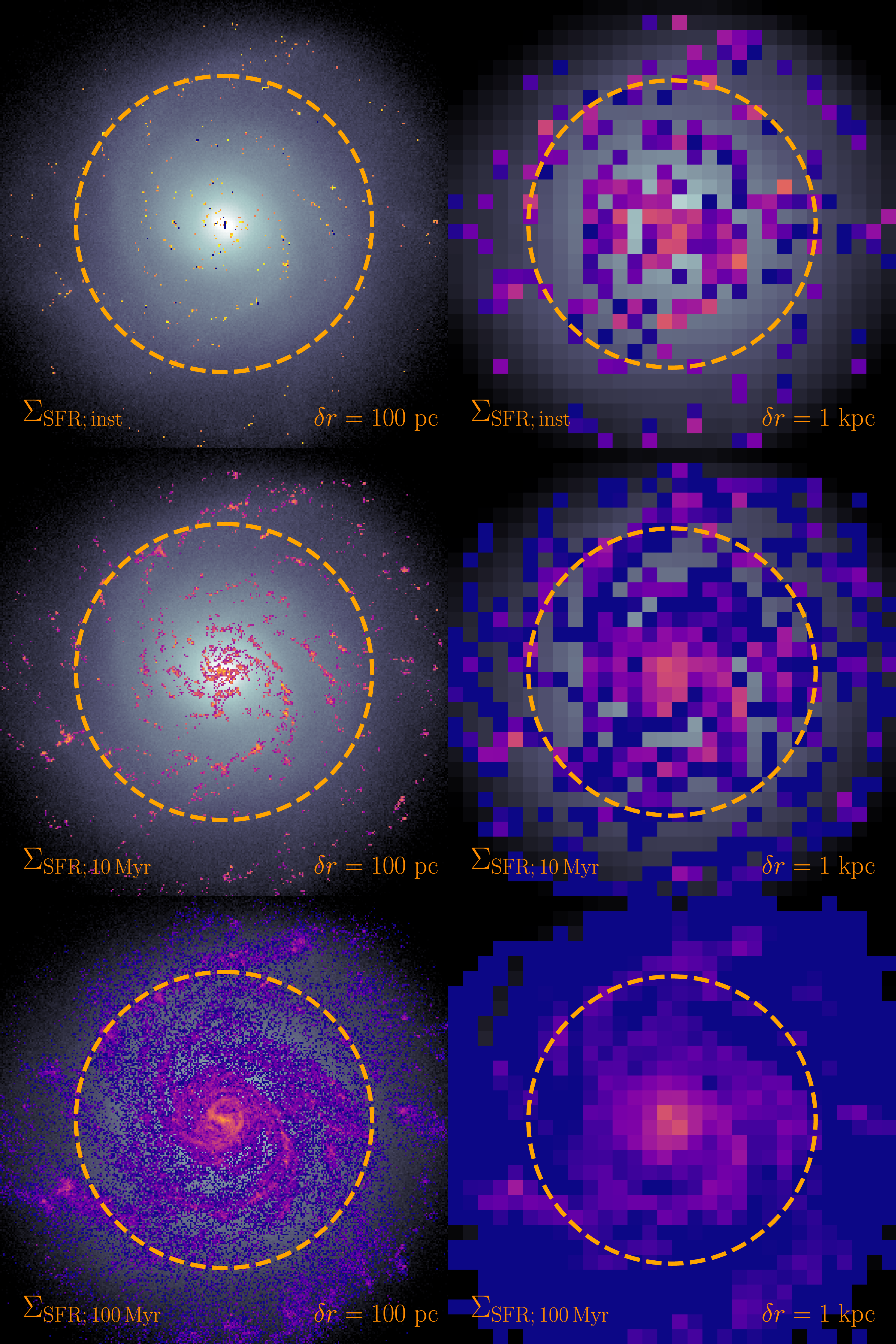}
\caption{Stellar mass surface density ($\Sigma_\star$; greyscale) maps overlaid with SFR surface density ($\Sigma_\mathrm{SFR}$; coloured pixels) maps of a sample simulated galaxy (\texttt{m12m}) at $100$ pc and $1$ kpc pixel scales (left and right hand panels, respectively). Top to bottom, the panels show the instantaneous gas SFR, $10$ Myr-averaged SFR, and $100$ Myr-averaged SFR in a $30\times 30$ kpc$^2$ field of view. The orange circle indicates $2\times r_\mathrm{half}$.}
\label{fig:FIRE_maps}
\end{figure}

\begin{figure*}
\centering
\includegraphics[trim=1.1cm 1.2cm 1.1cm 1.1cm, clip, width=0.8\textwidth]{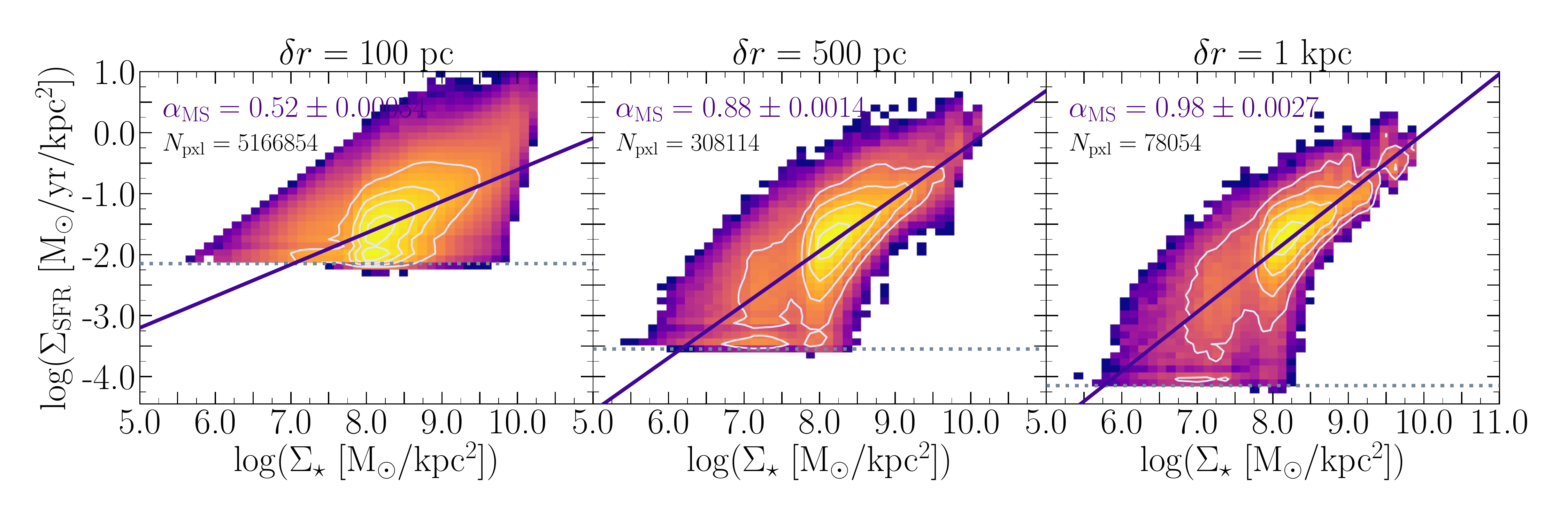}
\caption{The $100$ Myr-averaged rSFMS of the simulated galaxies using three pixel resolutions: $100$ pc, $500$ pc, and $1$ kpc (left to right). The solid lines show the best fit to the pixel data in the $\Sigma_\star$-$\Sigma_\mathrm{SFR}$ space, with the slope ($\alpha_\mathrm{MS}$) annotated in the top left corner. The contours include 20\%, 50\%, 70\%, and 90\% of the pixels. The minimum resolved $\Sigma_\mathrm{SFR}$ is indicated by the horizontal dotted lines: $\log (\Sigma_\mathrm{SFR}/\mathrm{M}_\odot \mathrm{yr}^{-1} \mathrm{kpc}^{-2})=-2.1$, $-3.5$, and $-4.1$ (left to right). An rSFMS is apparent at all spatial resolutions, and the relation flattens ($\alpha_\mathrm{MS}$ decreases) at higher spatial resolution. }
\label{fig:FIRE_MS}
\end{figure*}

% --------------------------------------------------------------------------------
% Spaxelization and SFMS fitting
% --------------------------------------------------------------------------------
Star formation rate surface density, and stellar mass surface density maps were generated for the FIRE-2 galaxies with $z \lesssim0.1$\footnote{The results presented in this work remain unchanged when using independent snapshots (every $\sim115$ Myr > $t_\mathrm{dyn}$).} following \citet{2018MNRAS.478.3653O}. Only particles within 15 kpc above/below the galactic disc contribute to the surface density maps\footnote{The particle selection is chosen to be consistent with \citet{2018MNRAS.478.3653O} who studies the Kennicutt-Schmidt relation in the same galaxy sample.}. The galaxies are deposited, face-on (along the stellar angular momentum axis), onto $30\times 30$ kpc$^2$ grids with resolutions ranging from $100$ pc to $5$ kpc. Note that we do not explore resolutions higher than $100$ pc, where we expect our star formation rate tracers to be affected by Poisson statistics (i.e. only a few young star particles contributing to a pixel). The SFR is measured using three SFR tracers: (1) `instantaneous' SFR, (2) $10$ Myr-averaged SFR, and (3) $100$ Myr-averaged SFR.
\footnote{Averaging time-scales of 10 and 100 Myr are often used
to crudely proxy observational SFR tracers such as recombination lines and FIR emission \citep[e.g.][]{2017MNRAS.466...88S}, but the time-scale probed by a specific tracer is sensitive to e.g. the detailed recent star formation history and how dust attenuation varies with age \citep[see table 1 of ][and the associated discussion]{2012ARA&A..50..531K}. We adopt these specific time-scales because they are commonly used in the literature, but our conclusions are insensitive to these choices.}
The time-averaged SFRs were calculated using the stellar particles' ages and correcting for mass-loss from stellar winds and stellar evolution. The instantaneous SFR refers to the SFR computed from the gas particles' densities and local free-fall timescales via the prescription stated above. Figure \ref{fig:FIRE_maps} shows a sample of the maps generate over a range of spatial resolutions and using different SFR time-scales.

Pixels within twice the stellar half-mass radius ($2\times r_\mathrm{half}$) with non-zero $\Sigma_{\rm SFR}$ are extracted from the maps, and the $\Sigma_\star - \Sigma_\mathrm{SFR}$ distribution is fit to investigate the effect of resolution and SFR tracer on the slope of the rSFMS ($\alpha_\mathrm{MS}$). We implement two fitting methods that are commonly used to fit the SFMS: ordinary least squares \citep[OLS; e.g. ][]{2018ApJ...857...17L}, and orthogonal distance regression \citep[ODR; e.g. ][]{2017ApJ...851L..24H}. While the absolute slope depends on the choice of fitting method (OLS, ODR, fitting medians, fitting modes), the qualitative trends shown in this work are robust to changes in the fitting method. Therefore, we adopt OLS to illustrate our results because it is more commonly used in the literature.

\section{Results}
\label{sec:results}
\noindent

% --------------------------------------------------------------------------------
% FIRE-2 results!
% --------------------------------------------------------------------------------
\subsection{The rSFMS in the FIRE-2 simulations}
\label{sec:results:fire}
We first investigate the existence of the rSFMS in the simulations. Figure \ref{fig:FIRE_MS} shows the distribution of pixels from the simulated galaxies in $\Sigma_\star - \Sigma_\mathrm{SFR}$ space at three example resolutions: $100$ pc, $500$ pc, and $1$ kpc. 

The rSFMS emerges in the simulations even though it is not explicitly prescribed in the sub-grid physics model. A similar rSFMS is also seen in individual galaxies. At resolutions of $1-2$ kpc the slope of the rSFMS of the simulated galaxies is broadly consistent with observations \citep[$\alpha_\mathrm{MS} \sim 0.62-1.09$ at $1-2$ kpc; ][]{2016ApJ...821L..26C, 2016A&A...590A..44G, 2017MNRAS.469.2806A, 2017ApJ...851L..24H, 2017MNRAS.466.1192M,  2018ApJ...857...17L}. 

\begin{figure}
\centering
\includegraphics[trim=1.2cm 1.17cm 0 1.15cm, clip, width=0.9\columnwidth]{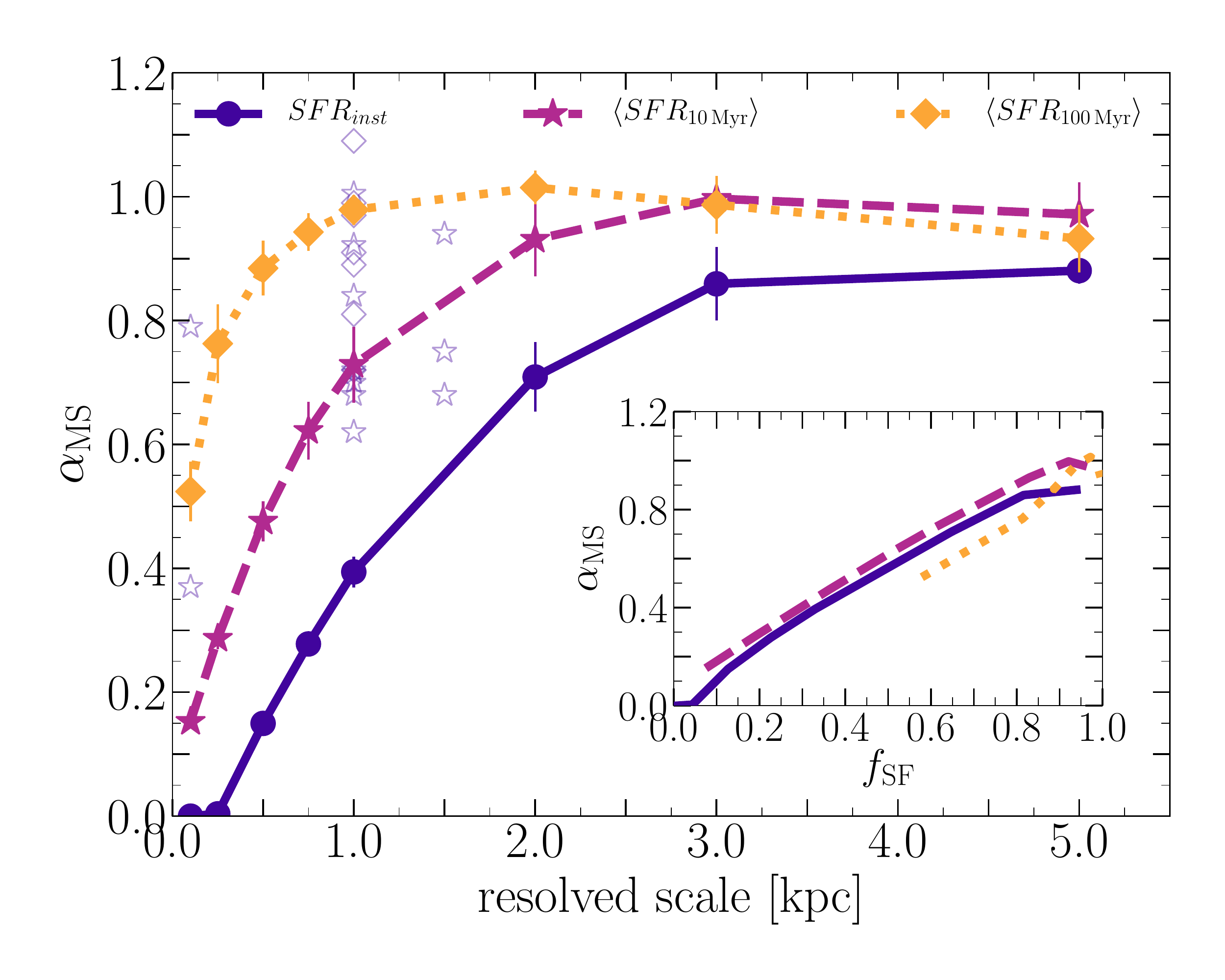}
\caption{The dependence of the rSFMS slope ($\alpha_\mathrm{MS}$) of the simulated galaxies on pixel resolution. The colours and line styles indicate the slopes derived from different SFR tracers (instantaneous, $10$ Myr, $100$ Myr), and the error bars are jackknife errors highlighting the galaxy-to-galaxy variations in our sample. At coarser resolutions, the rSFMS becomes steeper. The dependence of the slope on resolution is more pronounced for SFR tracers with shorter time-scales. For reference the open symbols indicate the slopes measured by various observational studies at low redshift ($z \lesssim 0.1$; see text for details). The inset shows the dependence of $\alpha_\mathrm{MS}$ on the filling factor of star-forming regions identified using the three SFR tracers. The slope depends strongly on the filling factor but is insensitive to the SFR tracer at a constant filling factor.}
\label{fig:FIRE_MS_res-dep}
\end{figure}

Figure \ref{fig:FIRE_MS_res-dep} shows the effect of pixel resolution on the measured rSFMS slope for our three SFR tracers (instantaneous, $10$ Myr--average, $100$ Myr--average). For reference, the open symbols show the measured $\alpha_\mathrm{MS}$ values from observational studies at low redshift \citep{2016ApJ...821L..26C, 2016A&A...590A..44G, 2017MNRAS.469.2806A, 2017ApJ...851L..24H, 2017MNRAS.466.1192M, 2018ApJ...857...17L, 2019MNRAS.484.5009E}. The different fitting methods and sample selections contribute to the scatter in the observed $\alpha_\mathrm{MS}$ \citep[e.g. ][]{2017ApJ...851L..24H, Lihwai, 2019MNRAS.488.1597V, Sara}. In the simulations, we find that as the spatial resolution increases, the rSFMS becomes shallower for all SFR tracers. The effect is more pronounced for the shorter-time-scale tracers. Additionally, at a given pixel scale, the slope depends on the SFR tracer. The shorter-time-scale tracers yield a shallower slope. We note that whilst the absolute value of the slope of the rSFMS and its dependence on resolution are sensitive to the details of the fitting method (such as choices of binning, weighting, SFR tracers, and threshold cuts), the generic steepening of $\alpha_\mathrm{MS}$ with lower resolution (or longer timescale SFR tracers) is robust to such choices. The dependence of $\alpha_\mathrm{MS}$ on spatial resolution and the time-scale of the SFR tracer represents an important caveat for comparisons between observational studies.

% --------------------------------------------------------------------------------
% Toy Model: The effect of clumpy star formation
% --------------------------------------------------------------------------------
\subsection{The effect of clumpy star formation}
\label{sec:results:toymodel}
\noindent
We propose that the reason for the dependence of $\alpha_\mathrm{MS}$ on spatial resolution and SFR tracer time-scale is the contrast between the clumpy distribution of star-forming regions and the relatively smooth underlying stellar mass distribution. As the spatial resolution decreases, a pixel's $\Sigma_\star$ remains roughly unchanged, while the effective contribution of an \textit{isolated} star-forming region decreases, thus causing a significant decrease in $\Sigma_\mathrm{SFR}$. This effect is less severe if there are more star-forming regions in a given pixel, since the higher filling factor mitigates the dilution effect of moving to lower resolution.  The time-scale of the SFR tracer has a similar effect -- as can be seen from Figure \ref{fig:FIRE_maps}, if SFRs are averaged over a longer time, they have a larger filling factor. Since the gas profile decreases radially, star-forming regions are sparsely separated in the outer regions of a given galaxy (cf. Figure \ref{fig:FIRE_maps}). Hence, the aforementioned effect of resolution and time-scales is more pronounced at lower $\Sigma_\star$ (which corresponds to larger radii, on average) thus causing a steeper $\Sigma_\star - \Sigma_\mathrm{SFR}$ correlation as the resolution decreases. The effect of resolution on the rSFMS slope is even more prominent for the shorter-time-scale SFR tracers, which trace smaller and `patchier' regions (see Figure \ref{fig:FIRE_maps}) and therefore are more severely affected by the change in pixel resolution.

\begin{figure}
\centering
\includegraphics[trim=1.cm 1.1cm 1.13cm 1.15cm, clip, width=0.9\columnwidth]{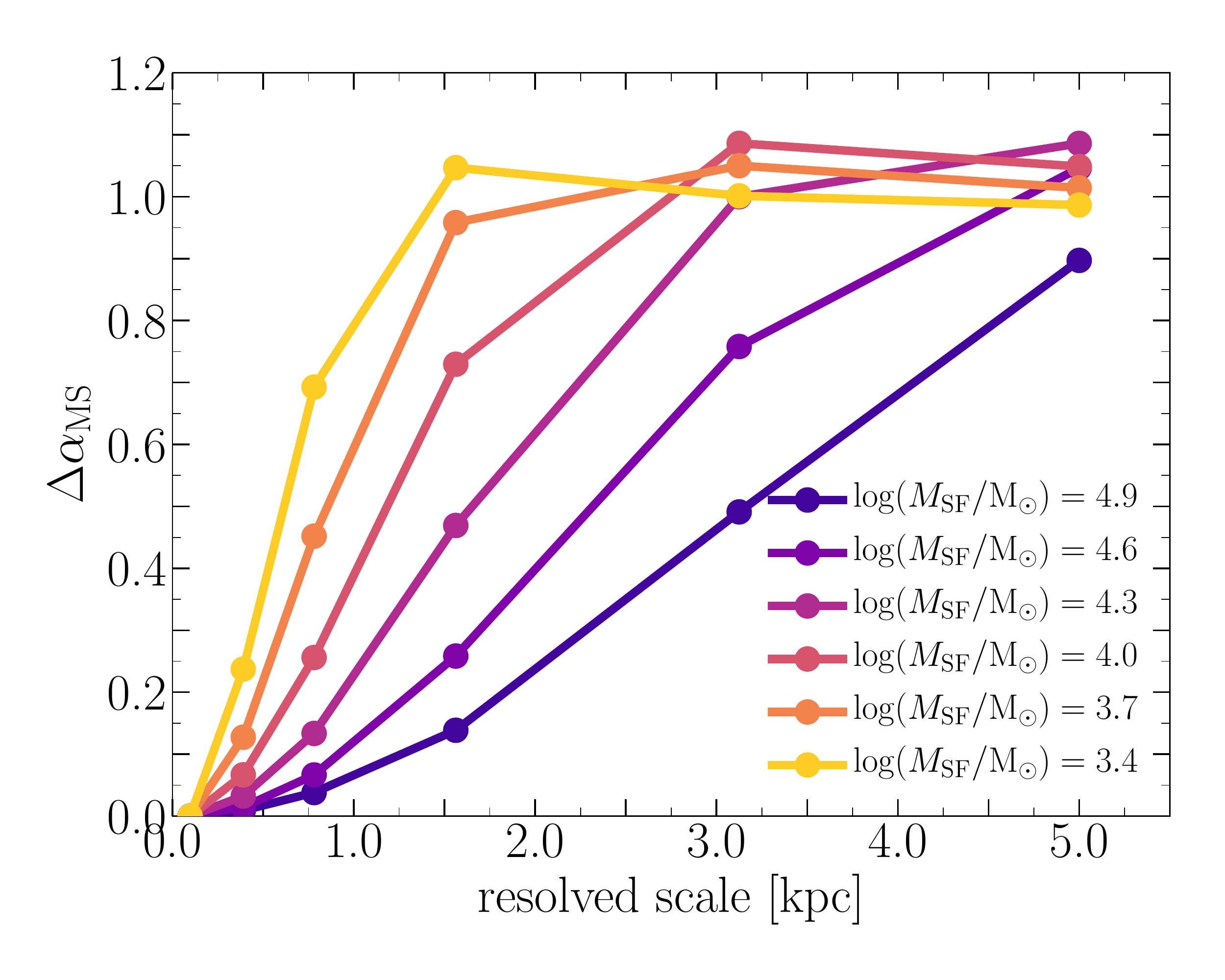}
\caption{Results from the toy model demonstrating the effect of pixel resolution on the slope of the rSFMS in the presence of clumpy star formation. The colours indicate different choices of $M_\mathrm{cut}$ in the mass distribution of star-forming clumps spawned in the synthetic galaxy sample. The legend reports the average clump mass ($M_\mathrm{SF}$) of the distribution. The dependence of $\alpha_{\rm MS}$ on spatial resolution is highly sensitive to the characteristic mass of star-forming regions.}
\label{fig:toy_model_MS_res-dep}
\end{figure}

In the scenario proposed above, the severity of the effect of pixel resolution (and SFR tracer time-scale) depends on the filling factor (and thus the total number, or equivalently the characteristic mass) of star-forming regions. The inset of Figure \ref{fig:FIRE_MS_res-dep} shows the dependence of the rSFMS slope on the fraction of pixels (within $2\times r_\mathrm{half}$) that host at least one star-forming region. The star formation filling factor correlates tightly with the rSFMS slope. As the star formation filling factor increases, the measured SFMS slope increases, thus supporting our hypothesis. Note that the SFMS slope exhibits a small dependence on the star formation time-scale at a given filling-factor, whereas $\alpha_\mathrm{MS}$ strongly depends on the filling factor.  
 
We further investigate the dependence of $\alpha_\mathrm{MS}$ on spatial resolution (and filling factor) with a toy model of galactic star formation. The toy model includes two components: (1) smooth synthetic stellar profiles and (2) clumpy star formation. We generate $10^4$ synthetic galaxies with a global $10$ Myr--averaged SFR of $1$ M$_\odot$/yr, appropriate for a $z \sim 0$ Milky Way-mass galaxy, and spawn them with star-forming regions as described below:
\begin{itemize}
    \item[$\rightarrow$]{\textbf{Stellar profile:} The surface density profiles are described by an azimuthally symmetric exponential disc profile: 
    \begin{equation*}
            \Sigma_\star (r) = \Sigma_{0} e^{-(r-R_0)/R_\mathrm{d}},
    \end{equation*}
    where $\Sigma_{0}= 6.1\times10^{7}$ M$_\odot /$kpc$^2$ is the stellar surface density at $R_0 = 8$ kpc, and the disc scale length is given by  $R_\mathrm{d} = 2.9$ kpc \citep{2016ApJ...816...42M}.} 
    
    \item[$\rightarrow$]{\textbf{Star-forming regions:}
    We assume an exponential gas profile parameterized by a scale of $3.75$ kpc \citep{2008A&A...487..951K}. We randomly spawn star-forming regions by sampling from a probability distribution following the gas surface density profile assuming a Kennicutt-Schmidt-like scaling (i.e. $P \propto \Sigma_{\rm gas}^{1.5}$; \citealt{1998ApJ...498..541K}) and deposit them onto a grid with a pixel scale of $20$ pc. The masses of star-forming regions are drawn probabilistically from a Schechter distribution \citep[e.g., ][]{2015MNRAS.452..246A}: $P\propto M^\eta \exp \left( -M/M_\mathrm{cut}\right)$ where $\eta = -2$ and $M_\mathrm{cut}$ is allowed to vary. Therefore, the number of star-forming regions is dictated by the choice of $M_\mathrm{cut}$ and the total SFR. Each distribution is characterised by an average clump mass: $M_\mathrm{SF}$. }
\end{itemize}

Note that the toy model is \textit{not} intended to be a \textit{fit} to the FIRE galaxies presented in Section \ref{sec:results:fire}. Instead, the parametric stellar profile and star-forming region distribution are chosen to broadly represent a synthetic Milky-Way profile. We emphasise that the results of the toy model do not depend on the scaling parameters, the total SFR, $\eta$, or $M_\mathrm{cut}$.

The synthetic galaxies are then degraded to lower resolutions, and the rSFMS is fitted following the method outlined above. Figure \ref{fig:toy_model_MS_res-dep} demonstrates the effect of pixel resolution on the measured rSFMS slope for different $M_\mathrm{SF}$. As the resolution decreases, the change in the rSFMS slope ($\Delta \alpha_\mathrm{MS} = \alpha_\mathrm{MS} - \alpha_\mathrm{MS; \, 100pc}$) is more drastic. Different choices of $M_\mathrm{SF}$ exhibit a distinct behaviour with varying resolution. The galaxies with a small $M_\mathrm{SF}$ (large star formation filling factor since a high number of regions is required to accumulate the fixed total SFR) show a less severe dependence on resolution beyond a critical resolved scale due to the decreased sensitivity to dilution effects when the pixel filling factor is high. For a \textit{large enough} pixel size, the underlying SFR field becomes statistically uniform, thus causing the rSFMS slope to exhibit a shallower dependence on resolution. The dependence of $\alpha_\mathrm{MS}$ on resolution remains unchanged if the clump mass was fixed to $M_\mathrm{SF}$. 

\section{Summary \& Discussion}
\label{sec:conclusions}
\noindent
In this Letter, we present the first analysis of the rSFMS of Milky Way-mass galaxies in the FIRE-2 simulations. We have demonstrated that the SFR tracer time-scales and pixel resolution can both have a significant impact on the derived slope of the rSFMS. We have shown that this effect is due to the clumpy nature of star formation; hence the measured slope of the rSFMS depends directly on the filling factor of star-forming regions, and the physical properties (i.e. mass) of star-forming clumps.

The dependence of the rSFMS slope on resolution and SFR tracer can be used to constrain properties (e.g. mass) of star-forming clumps (see Figure \ref{fig:toy_model_MS_res-dep}). Using a more \textit{realistic} distribution of star-forming clump masses (e.g. a radially dependent Schechter function or power-law), the toy model presented in this work ($\S$\ref{sec:results:toymodel}) could potentially be used to infer observed average star-forming clump masses from the dependence of $\alpha_\mathrm{MS}$ on resolution. Additionally, the variation in the rSFMS slope may hint that the rSFMS may not be a fundamental diagnostic of star formation. Instead, understanding the scales at which the rSFMS deviates from the global and kpc-scale relations may shed more light on star formation on galactic scales.

Although the work presented here focuses on the rSFMS, the same slope dependence on resolution will be seen for other relationships that combine clumpy properties (here, $\Sigma_\mathrm{SFR}$) with more smoothly varying ones (here, $\Sigma_\star$). Conversely, relationships that use quantities that probe similar scales (such as the Kennucutt-Schmidt law) should not be strongly affected by resolution. Indeed, \citet{2018MNRAS.478.3653O} report that the Kennicutt-Schmidt law does not depend on resolution over a similar resolution range as studied in this work.

The work presented in this Letter has notable implications for IFU surveys. Particularly, comparisons between surveys can be undermined by several effects, including (1) different pixel resolutions and (2) different SFR tracers, which are sensitive to different time-scales (and thus physical scales).
Moreover, by progressively degrading high-resolution IFU data to coarser resolutions and analyzing the dependence of the rSFMS slope on spatial resolution, the characteristic mass of star-forming regions can be inferred using toy models similar to the one presented here.

\section*{Acknowledgements}
The authors thank the anonymous referee for their helpful comments which improved the presentation of this work. The authors thank Connor Bottrell, Greg Bryan, John Forbes, Shy Genel, Li-Hwai Lin, Nic Loewen, Ari Maller, Hsi-An Pan, David Patton, Rachel Somerville, Mallory Thorp, and Joanna Woo for their insightful comments and helpful discussions. MHH acknowledges the receipt of a Vanier Canada Graduate Scholarship. SLE acknowledges the receipt of an NSERC Discovery Grant. The data used in this work were, in part, hosted on facilities supported by the Scientific Computing Core at the Flatiron Institute, a division of the Simons Foundation.

%%%%%%%%%%%%%%%%%%%%%%%%%%%%%%%%%%%%%%%%%%%%%%%%%%


\bibliographystyle{mnras}
\bibliography{bibliography} % if your bibtex file is called example.bib

%%%%%%%%%%%%%%%%%%%%%%%%%%%%%%%%%%%%%%%%%%%%%%%%%%

% Don't change these lines
\bsp	% typesetting comment
\label{lastpage}
\end{document}